\documentclass[aps,twocolumn,prd,showpacs,nofootinbib]{revtex4}
\usepackage{amsmath}
\usepackage{graphicx}
\usepackage{subfigure}
\usepackage{dcolumn}
\usepackage{bm}
\usepackage{amssymb}
\usepackage{latexsym}
\usepackage{psfrag}

\def\setC{\mathbb{C}}

\def\setR{\mathbb{R}}

\newcommand{\dd}{\mathrm{d}}

\newcommand{\lsim}{\lesssim}
\newcommand{\ie}{\textsl{i.e.~}}

\newcommand{\mP}{m_{_{\mathrm Pl}}}
\newcommand{\mC}{m_{_{\mathrm C}}}
\newcommand{\GReCO}{${\cal G}\setR\varepsilon\setC{\cal O}$}

\def\spose#1{\hbox to 0pt{#1\hss}}
\def\lta{\mathrel{\spose{\lower 3pt\hbox{$\mathchar"218$}}
     \raise 2.0pt\hbox{$\mathchar"13C$}}}
\def\gta{\mathrel{\spose{\lower 3pt\hbox{$\mathchar"218$}}
     \raise 2.0pt\hbox{$\mathchar"13E$}}}

\begin{document}

\preprint{BROWN-HET-1400}
\preprint{MCGILL-10-04}

\title{Back-Reaction and the Trans-Planckian Problem of Inflation
Revisited}

\author{Robert H. Brandenberger} \email[Email:]{rhb@het.brown.edu}
\affiliation{Department of Physics, McGill University, Montr\'eal,
Qu\'ebec, H3A 2T8, Canada,\\ Perimeter Institute for Theoretical
Physics, Waterloo, N2J 2W9, Canada, and \\ Department of Physics,
Brown University, Providence, RI 02912, USA}

\author{J\'er\^ome Martin}\email[Email:]{jmartin@iap.fr}
\affiliation{Institut d'Astrophysique de Paris, \GReCO, 98bis boulevard Arago,
75014 Paris, France}

\date{\today}

\begin{abstract}
It has recently been suggested that Planck scale physics may effect
the evolution of cosmological fluctuations in the early stages of
cosmological inflation in a non-trivial way, leading to an excited
state for modes whose wavelength is super-Planck but sub-Hubble. In
this case, the issue of how this excited state back-reacts on the
background space-time arises. In fact, it has been suggested that such
back-reaction effects may lead to tight constraints on the magnitude
of possible deviations from the usual predictions of inflation. In
this note we discuss some subtle aspects of this back-reaction issue
and point out that rather than preventing inflation, the back-reaction
of ultraviolet fluctuations may simply lead to a renormalization of
the cosmological constant driving inflation.
\end{abstract}

\pacs{98.80.Cq.}

\maketitle

\section{Introduction}

The most important success of the inflationary Universe scenario
\cite{Guth} is that it provides a causal mechanism for the origin of
the observed density fluctuations and microwave background
anisotropies~\cite{Mukhanov} (see also Refs.~\cite{Press,Lukash}). Key
to this success is the fact that the physical wavelength corresponding
to a fixed comoving scale is exponentially stretched during the period
of inflation. Thus, provided that the period of inflation lasts
sufficiently long, fluctuations on scales of cosmological interest
today originate on sub-Hubble scales during inflation. Since inflation
red-shifts all initial classical fluctuations, it is reasonable to
assume that matter starts out in a quantum vacuum state (in the frame
set by the background cosmology, see e.g. Ref.~\cite{RHB84} for a
discussion). Each fluctuation mode thus starts out in its vacuum state
at the time that the initial conditions are set up (e.g. the beginning
of the period of inflation), it undergoes quantum vacuum oscillations
while the wavelength is smaller than the Hubble radius, but freezes
out when the wavelength equals the Hubble radius (see
e.g. Ref.~\cite{MFB} for comprehensive reviews of the theory of
cosmological fluctuations). Subsequently, the quantum state of the
fluctuations undergoes squeezing on super-Hubble scales, and re-enters
the Hubble radius during the post-inflationary Friedman-Lema\^\i
tre-Robertson-Walker (FLRW) phase as a highly squeezed and effectively
classical state (see e.g. Ref.~\cite{Polarski} for a discussion of the
classicalization of the state).

\par

However, as first pointed out in Ref.~\cite{RHBrev}, this success of
inflationary cosmology leads to an important conceptual problem, the
{\it trans-Planckian problem}. Since the period of inflation in
typical scalar-field-driven inflationary models is very long (see
e.g. Ref.~\cite{Lindebook} for a review), the scales of cosmological
interest today are not only sub-Hubble, but in fact sub-Planck at the
beginning of inflation. Thus, the formalism used to calculate the
evolution of fluctuations is in fact not justified. It is possible
that the unknown trans-Planckian physics will lead to an evolution of
the fluctuations on sub-Planckian scales which from the point of view
of free scalar field theory coupled to General Relativity looks
non-adiabatic.  In Ref.~\cite{MB1} (see also Ref.~\cite{Niemeyer}),
toy models for such an evolution were constructed making use of
modified dispersion relations which were assumed to describe the
physics on sub-Planckian scales. Since the time interval spent in the
trans-Planckian domain may depend on the wavelength, such models may
lead to changes in the spectral index of the fluctuations.

\par

Subsequently, other approaches to the trans-Planckian problem were
suggested, e.g. analysis based on space-space
non-commutativity~\cite{Columbia}, space-time
non-commutativity~\cite{Ho}, minimal trans-Planckian assumptions
(starting each mode in some vacuum state at the time when its
wavelength equals the Planck length)~\cite{Ulf,Ven}. These analysis
typically give that trans-Planckian corrections to the predictions for
cosmological fluctuations are proportional to $(H_{\rm inf}/\mC)^n$
where $H_{\rm inf}$ is the Hubble parameter during inflation, $\mC$ a
new scale at which non standard physical effects show up and $n$ a
number which depends on the initial state assumed at the time of
``creation''. On the other hand, analysis based on modified dispersion
relations give a correction proportional to the time spent by the
physical modes in the region where adiabaticity is violated, see
e.g.~\cite{MB2} for a recent review.

\par

Tanaka~\cite{Tanaka} and Starobinsky~\cite{Starob} (see also
Ref.~\cite{later}) have, however, raised an important concern
regarding the possible amplitude of trans-Planckian corrections (see
also Ref.~\cite{TS} for unrelated concerns): if trans-Planckian
physics leads to an excited state for fluctuation modes on sub-Hubble
but super-Planck scale during the period of inflation, the
back-reaction of these excitations on the background must be
considered. In the case where the trans-Planckian effects are modeled
by a modified dispersion relation, a simple estimate of the energy
density carried in these ultraviolet modes
\begin{equation} 
\label{estimate} 
\langle \rho \rangle _{_{\rm UV}} = \int_{k_{\rm phys} = H_{\rm inf}}
^{k_{\rm phys} = \mC} {\rm d}^3k_{\rm phys} \omega_{\rm phys}(k_{\rm
phys}) n_{k_{\rm phys}}  \, ,
\end{equation}
where $k_{\rm phys}$ is the physical wavenumber, $n_{k_{\rm phys}}$ is
the occupation number, and $\omega_{\rm phys}$ is the frequency of the
mode, leads to the conclusion that $\langle \rho \rangle _{_{\rm UV}}$
will exceed the background density unless $n_{k_{\rm phys}}$ is
smaller than $(H_{\rm inf}/m_{_{\rm Pl}})^2$ [see also
Eq.~(\ref{consbeta}) above], thus constraining the possible effects of
trans-Planckian physics on the spectrum of fluctuations (in the above,
we have assumed that the ultraviolet cutoff is the usual Planck scale
$m_{_{\rm Pl}}$ which is not mandatory at all).

\par

In this article, we point out some subtleties with the above
back-reaction argument which may change the conclusions dramatically.
If we assume, as is conventionally done in analyzing quantum fields in
curved space-time, that the ultraviolet cutoff scale is
time-independent in terms of physical length, then in an exponentially
expanding background geometry, the contribution of ultraviolet
(i.e. sub-Hubble) modes to the energy density is constant in time, as,
in fact, follows directly from the time translation invariance of the
physics. Moreover, the corresponding equation of state, due to the
fact that the dispersion relation is modified, can strongly differ
from that of ultra-relativistic particles and, as we demonstrate
below, tends to that of the vacuum. Hence, our main conclusion is
that, instead of preventing inflation, the ultraviolet modes may in
fact simply renormalize the value of the cosmological constant driving
inflation.

\par

This article is organized as follows. In the next section, Sec.~II, we
describe the arguments that have been put forward to claim that there
is a back-reaction problem and we criticize them. Then, in Sec.~III, we
present an explicit calculation of the equation of state of a scalar
field with a modified dispersion relation. We show that the
ultraviolet modes possess an equation of state which is almost that of
a cosmological constant. Finally, in Sec.~IV, we point out problems
with our approach, indicate directions for further investigations and
present our general conclusions.

\section{The Back-Reaction Problem}

As explained in the introduction, it is possible that the
trans-Planckian effects affect the standard inflationary predictions.
In this paper, for the sake of illustration, we model physics at very
short scales by a non-linear dispersion relation $\omega _{\rm
phys}(k_{\rm phys})$. For wave-numbers such that $k_{\rm phys }\ll
\mC$ where $\mC$ is a new scale at which non-standard physical effects
show up, the dispersion relation is linear for obvious
phenomenological reasons. On the contrary, for modes such that $k_{\rm
phys }\gg \mC$, the shape of $\omega _{\rm phys}(k_{\rm phys}) $ is a
priori unknown. It is has been shown that a non-adiabatic evolution of
the mode function in the trans-Planckian region necessarily implies a
modification of the inflationary predictions, in particular a
modification of the power spectrum. To be more precise, in cosmology,
the dispersion relation becomes time-dependent and equal to
$\omega=a\omega _{\rm phys}(k/a)$ ($\omega$ and $k$ denote comoving
frequency and wavenumber, respectively).  Then, the
Wentzel-Kramers-Brillouin (WKB) approximation is satisfied provided
that $\vert Q/\omega ^2\vert \ll 1$, where the quantity $Q$ is defined
by $Q=3(\omega ')^2/(4\omega ^2)-\omega ''/(2\omega )$ (a prime stands
for the derivative with respect to conformal time), see also
Ref.~\cite{MS}. If the previous condition is worked out, then one sees
that the WKB approximation is violated if $\omega _{\rm phys}<H_{\rm
inf}$ and that corrections to the standard result can occur in this
case.

\par

This conclusion has been criticized in Refs.~\cite{Tanaka,Starob} and
many reasons why a modification of the inflationary power spectrum
would be unlikely have been provided in these articles. In the
following, we will examine each of them.

\par

In Ref.~\cite{Starob}, it has been claimed that if $\omega _{\rm
phys}(k_{\rm phys})$ is such that the WKB approximation is violated
for $k_{\rm phys}>\mC$ then there are no preferred initial
conditions. This is certainly correct for the class of dispersion
relations considered in Ref.~\cite{Laura}, as discussed in
Ref.~\cite{Lubo}, but not true in general. An explicit counter-example
has been provided in Ref.~\cite{Lubo} and is studied in the present
paper. The corresponding dispersion relation is sketched in
Fig.~\ref{disp}. On this plot, one notices that in the region where
$k_{\rm phys} > \mC$ there is an interval of finite range, namely
$k_{\rm phys} \in [\Lambda _1,\Lambda _2]$, in which the WKB
approximation is violated (\ie $\omega_{\rm phys}<H_{\rm inf}$).
Important for our analysis is also the fact that for $k_{\rm
phys}\rightarrow +\infty $ the adiabatic approximation is restored,
which follows since we have $\omega _{\rm phys} > H_{\rm inf}$.  In
this latter regime, the adiabatic vacuum is obviously the preferred
initial state.

\par

The other arguments involve the calculation of the stress-energy
tensor and are as follows. In the standard scenario, the initial
conditions are fixed for all wavenumbers at some initial time. If the
number of e-foldings of inflation is greater than about 70, the
physical wavelength of modes which are currently probed in cosmic
microwave experiments is smaller than the Planck length at the
beginning of inflation.  One usually assumes that the evolution starts
out from the adiabatic (Bunch-Davis) vacuum. If the evolution is
non-adiabatic in the region $k_{\rm phys}>\mC$, then the state in the
region $H_{\rm inf}<k_{\rm phys}<\mC$ will differ from the usual
adiabatic vacuum. Therefore, if one concentrates only on what happens
in the region $H_{\rm inf}<k_{\rm phys}<\mC$, the trans-Planckian
effects boil down to a modification of the initial conditions. This
last argument has been used in Refs.~\cite{Tanaka, Starob} as
follows. Roughly speaking, a non-vacuum state means a non-vanishing
energy density and there is now the danger that this dominates over
the energy density of the inflationary background which is $m_{_{\rm
Pl}}^2H_{\rm inf}^2$. According to Refs.~\cite{Tanaka, Starob}, this
is actually what happens unless the level of excitation of the initial
state compared to the adiabatic vacuum is very small, leading to
unmeasurably small trans-Planckian effects of the spectrum of
fluctuations. Since observational evidence seems to indicate that
inflation is the correct theory of the very early universe,
Refs.~\cite{Tanaka, Starob} conclude that trans-Planckian effects of
significant importance are in fact not possible.

\par

In order to understand the above argument in more details, let us be
more accurate about what has actually been done in
Refs.~\cite{Tanaka,Starob}. It is well-known that cosmological
perturbations (density fluctuations and gravitational waves) can, in
some contexts, be viewed as a free scalar field $\varphi (\eta ,{\bf
x})$ on a time-dependent background space-time. In the case of
gravitational waves, the correspondence is exact, for scalar metric
fluctuations (density perturbations), the correspondence is only exact
if the equation of state of the background is time-independent. In the
general case, the squeezing factor for the density fluctuations is given
not by the FLRW scale factor $a(t)$, but by a function $z(t)$ which
depends both on the background geometry and the background matter -
for details see e.g. \cite{MFB}. Then, the corresponding energy
density and pressure are given by the mean values of the stress-energy
tensor $\langle T_{\mu \nu}\rangle $. In an excited state
characterized by the mode distribution function $n=n(k)$, one has
\begin{widetext}
\begin{eqnarray}
\langle \rho \rangle &=&
\frac{1}{4\pi ^2 a^4}\int _0^{+\infty } {\rm d}k k^2
\left[\frac12+n(k)\right]\times 
\left[a^2\left|\left(\frac{\mu_k}{a}\right)'\right|^2
     +k^2\left|\mu_k\right|^2\right]\, ,
\\
\langle p \rangle &=&
\frac{1}{4\pi ^2 a^4}\int _0^{+\infty } {\rm d}k k^2
\left[\frac12+n(k)\right]\times 
\left[a^2\left|\left(\frac{\mu_k}{a}\right)'\right|^2
     -\frac{k^2}{3}\left|\mu_k\right|^2\right]\, ,
\end{eqnarray}
\end{widetext}
where $\mu _k$ is the rescaled Fourier amplitude, \ie $\mu _k\equiv
a(\eta ) \varphi _k(\eta )$ and normalized such that $\mu _k\simeq
1/\sqrt{2k}$. In the above, $\eta$ denotes conformal time, and a prime
the derivative with respect to $\eta$. 

\par

In the above expressions, the terms proportional to the factor $1/2$
are divergent in the ultra-violet regime, i.e. $k\rightarrow +\infty
$, and represent the quantum vacuum contribution.  This means that, in
order to give sense to the above expressions, the stress energy tensor
should be first properly renormalized, for instance by adiabatic
regularization~\cite{PF}. In this paper, we will simply subtract the
contribution of the quantum vacuum energy since this is what has been
done in Refs.~\cite{Tanaka, Starob}. Another justification is that we
are in fact mainly interested in the terms proportional to $n(k)$
which describe the contributions originating from the excited quanta.

\par

In Refs.~\cite{Tanaka,Starob}, considerations have been restricted to
physical modes such that $H_{\rm inf}<k_{\rm phys}<\mC$. In this
region, the dispersion relation is linear and, since the WKB
approximation is satisfied, the mode function can be written as
\begin{equation}
\label{wkbstandard}
\mu _k=\frac{\alpha _k}{\sqrt{2k}}{\rm e}^{-ik\eta }
+\frac{\beta _k}{\sqrt{2k}}{\rm e}^{+ik\eta }\, ,
\end{equation}
where $\vert \alpha _k\vert ^2-\vert \beta _k\vert ^2=1$. Inserting
this mode function into the vacuum expressions of the energy density
and pressure, one finds
\begin{eqnarray}
\langle \rho \rangle _{_{\rm UV}} &=&
\frac{1}{2\pi ^2 a^4}\int _{aH_{\rm inf}}^{a\mC} \frac{{\rm d}k}{k}
k^4\vert \beta _k\vert ^2\, , \label{rhouv}
\\
\langle p \rangle _{_{\rm UV}} &=&
\frac{1}{2\pi ^2 a^4}\frac{1}{3}\int _{aH_{\rm inf}}^{a\mC} \frac{{\rm
d}k}{k}
k^4\vert \beta _k\vert ^2\, , \label{uv2}
\end{eqnarray} 
We see that the coefficient $\vert \beta _k\vert ^2$ represents the
number of particles, $n(k) =\vert \beta _k\vert ^2$. This describes
the modification of the standard initial conditions due to the
trans-Planckian effects (let us remind that the usual adiabatic
initial conditions correspond to $\alpha _k=1$ and $\beta
_k=0$). Then, a back-of-the-envelope calculation shows that $\langle
\rho \rangle _{_{\rm UV}}\simeq \mC^4\vert \beta _k\vert ^2$ where we
have used the fact that, in de Sitter space-time, the coefficient
$\beta _k$ is scale-independent (time translation invariance). If we
require that the energy density of the test scalar field be smaller
than the background density, then it follows that
\begin{equation}
\label{consbeta}
\vert \beta _k\vert ^2<\frac{m_{_{\rm Pl}}^2H_{\rm inf}^2}{\mC^4}\, .
\end{equation}
A similar expression has been obtained in Ref.~\cite{Starob}, with
$\mC=m_{_{\rm Pl}}$ and, crucially, $H_{\rm inf}$ replaced with $H_0$,
the present value of the Hubble parameter. It is clear that the
constraint on $\beta _k$ is completely different (and much more
difficult to satisfy) if one uses $H_0\simeq 10^{-61}m_{_{\rm Pl}}$ in
the above equation rather than $H_{\rm inf}\simeq 10^{-5}m_{_{\rm
Pl}}$ (we also notice that $\mC$ needs not be the Planck mass). The
reason for this difference is again (see above the discussion of the
preferred initial conditions) that, in Ref.~\cite{Starob}, it was
assumed that violation of the WKB approximation in the trans-Planckian
region necessarily implies that $\omega (k) \rightarrow 0$ as
$k\rightarrow +\infty$, as for the dispersion relation envisaged in
Ref.~\cite{Laura}. In this case, the adiabatic condition is violated
today for a range of trans-Planckian modes, and one should indeed
replace $H_{\rm inf}$ by $H_0$ as done in Ref.~\cite{Starob}. This
results in a very strong constraint on $\beta_k$. This was in fact the
essence of the criticism made in Ref.~\cite{Lubo} against the
dispersion relation considered in Ref.~\cite{Laura}. However, again,
the argument does not apply for dispersion relations of the type shown
in Fig.~\ref{disp} and, therefore, is not true in general. The reason
can be very easily understood from Fig.~\ref{disp}. Since, after
inflation, the Hubble parameter decreases and since, at some point,
its value becomes smaller than the minimum of the dispersion relation,
the adiabatic condition is restored at late times for all modes, and
particle production stops. Therefore, in this case, the calculation
should be done with $H_{\rm inf}$ and not with $H_0$ as done in
Ref.~\cite{Starob}. In Refs.~\cite{Lubo, MB2}, it has been shown that
there is a window for which the modification of the power spectrum is
not completely negligible and for which there is no back-reaction
problem, see for instance the discussion after Eq.~(68) in
Ref.~\cite{Lubo}.

\par

The other arguments presented against possible trans-Planckian
modifications of the inflationary power spectrum involve the
calculation of the equation of state (which is characterized by
the parameter $\omega _{\rm st} = p / \rho$). The main argument is that the
``dangerous'' created particles behave as a radiation field (or as
ultra-relativistic particles) and, therefore, that their energy
density scales as $a^{-4}$ and not as the vacuum, see
Ref.~\cite{Tanaka}. We now explain why this line of reasoning is
problematic.

\par

{F}irst of all, using the stress-energy tensor of a test scalar field
is questionable since one wants in fact to calculate the equation of
state of the cosmological perturbations, not that of a test scalar
field.  It has been shown in Refs.~\cite{MAB,ABM} that the equation of
state of the effective stress-energy tensor of the cosmological
perturbations differs on super-Hubble scales (which is a region in
which the adiabatic condition is not valid and thus has
similarities to the trans-Planckian interval where the adiabatic
condition is violated) from the equation of state of a test field. It
is $\omega _{\rm st} = -1$ instead of $\omega _{\rm st} = -1/3$. The
difference is clearly of utmost importance in the present context. In
fact, what should be done is to calculate the stress-energy
tensor of cosmological perturbations (which is second order in the
perturbed metric) in the case where the dispersion relation is
modified. To our knowledge, this calculation has never been performed,
and is very complicated. Therefore, this is beyond the scope of the
present article.

\par

Keeping the previous point in mind, let us come back to the
calculation of the stress-energy tensor of a test scalar field. As
already mentioned in the introduction [and demonstrated explicitly 
in Eq.~(\ref{enerphys})], $\langle \rho \rangle _{_{\rm UV}}$ is in fact
constant and does not scale as $a^{-4}$ despite the fact that $p/\rho
=1/3$. The non-conservation of the energy-momentum tensor 
can be understood as follows. Using the
expression of the energy density, it is easy to establish that
\begin{eqnarray}
\label{nonconserv}
\frac{1}{a^4}\frac{{\rm d}}{{\rm d}t}\left(a^4\langle \rho \rangle
_{_{\rm UV}}\right) &=& \frac{H_{\rm inf}}{2\pi ^2}\biggl[\mC^4\vert 
\beta _{k=a\mC}\vert ^2
\nonumber \\
& & -H_{\rm inf}^4\vert \beta _{k=aH_{\rm inf}}
\vert ^2\biggr]\, .
\end{eqnarray}
The two terms in the right-hand-side of the
above expression, responsible for the non-conservation, originate from
the time-dependent limits of integration in Eq.~(\ref{rhouv}). The
first one comes from the upper limit while the second one originates
from the lower limit. The corresponding physical interpretation is
clear: due to the expansion of the background, there is a flow of
modes coming from the trans-Planckian region and entering the region
$H_{\rm inf}<k_{\rm phys}<\mC$ and there is also a flow of modes
leaving the region $H_{\rm inf}<k_{\rm phys}<\mC$ while they are
becoming super-Hubble modes. Eq.~(\ref{nonconserv}) is similar to
Eq.~(4) of Ref.~\cite{Starob}. The only difference is the absence in
Ref.~\cite{Starob} of the second term on the right-hand-side
(describing the outgoing flow of modes). This term is necessarily
present because the integral in Eq.~(\ref{rhouv}) cannot be computed
with a vanishing lower integral since, for $k_{\rm phys}<H_{\rm inf}$,
the mode function is no longer given by Eq.~(\ref{wkbstandard}) but
rather by $\mu _k\simeq a(\eta )$. However, the second term in 
(\ref{nonconserv}) is
clearly very small in comparison with the first one and, therefore,
can be safely neglected.

\par

{F}inally, maybe the most important reason why calculating the
back-reaction can be more subtle than previously thought is the
following. The conclusion that $p/\rho=1/3$ is in fact obtained from
an inconsistent procedure since it does not take into account the fact
that the dispersion relation is modified [let us remind that the
previous considerations are based on Eqs.~(\ref{rhouv})
and~(\ref{uv2}) that have been obtained under the assumption that
$\omega _{\rm phys}=k_{\rm phys}$]. In the following, we shall study
the equation of state of the ultraviolet terms (\ref{rhouv}) and
(\ref{uv2}) in a toy model for trans-Planckian physics in which we can
describe the excitation of the mode functions in the far ultraviolet
range from well-defined vacuum initial conditions in a mathematically
consistent way. We shall show that the fact that the dispersion
relation is modified can change the equation of state, a conclusion
also reached in Ref.~\cite{Jacobson} in a slightly different context.

\par

To summarize, the calculation of the back-reaction must be performed
with the trans-Planckian corrections taken into account (\ie in the
present context with a modified dispersion relation). It is clearly
inconsistent to calculate the modified power spectrum with the
trans-Planckian corrections on one hand and, on the other hand, to
evaluate the corresponding back-reaction without these
corrections. This can change, in a crucial way, the calculation of the
energy density and/or the equation of state.

\section{The toy model}

\subsection{Description of the Dispersion Relation}

We model the trans-Planckian effect by means of the following
non-standard dispersion relation \cite{Lubo}
\begin{equation}
\omega ^2_{\rm phys}\left(k_{\rm phys}\right) \, = \,
k^2_{\rm phys}-2b_{11}k^4_{\rm phys}+2b_{12}k^6_{\rm
phys} \, . 
\end{equation}
This dispersion relation is chosen such that the modes
evolve adiabatically for extremely high wavenumbers, but, given
an appropriate choice of the constants $b_{11}$ and $b_{12}$,
there is an intermediate region of wavenumbers in which the mode
evolution is not adiabatic. We will start the modes in their
adiabatic vacuum in the extreme ultraviolet and calculate how
they are excited during the phase in which the evolution violates
the adiabaticity condition. 

\begin{figure*}
\includegraphics[width=.93\textwidth,height=.55\textwidth]{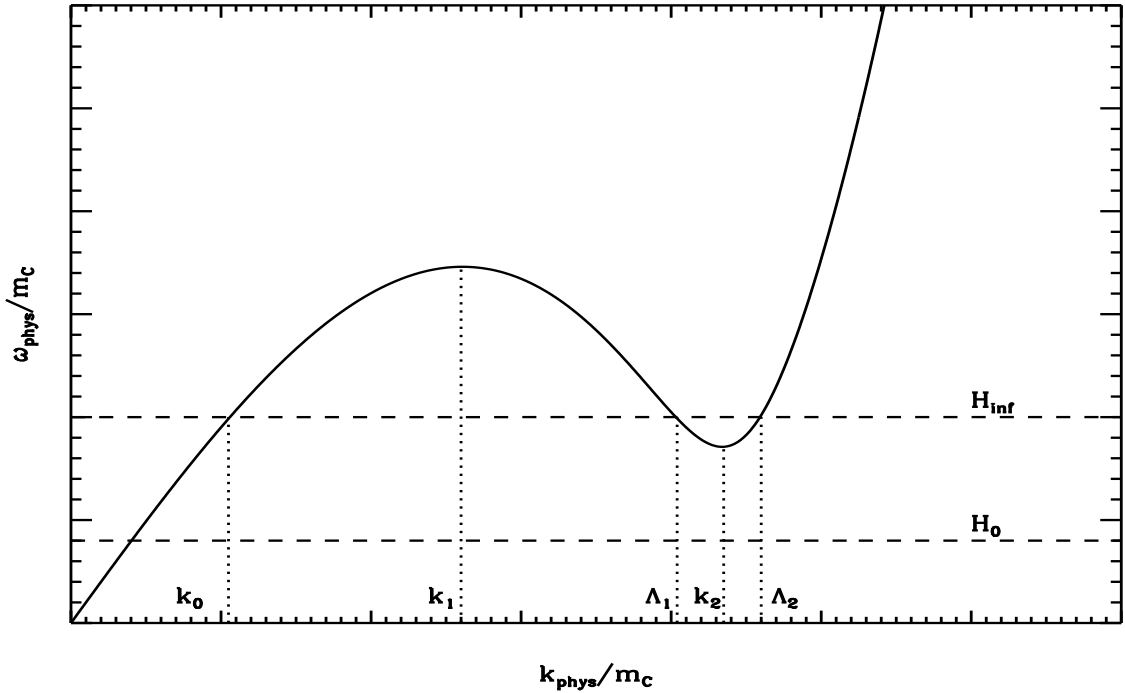}
\caption{Sketch of the dispersion used for the study of the toy
model. The five scales $k_0$, $k_1$, $k_2$, $\Lambda _1$ and $\Lambda
_2$ are defined in the text. $H_{\rm inf}$ is the Hubble parameter
during inflation while $H_0$ is the Hubble parameter today. Clearly,
this is not a scaled figure since $H_0$ should be much smaller than
is represented on the plot.} 
\label{disp}
\end{figure*}

If we introduce the new dimensionless coefficients $\alpha $
and $\beta $ such that $\alpha \equiv 2b_{11}m_{_{\rm C}}^2$ and
$\beta \equiv 2b_{12}m_{_{\rm C}}^4$, where $m_{_{\rm C}}$ is a new
free energy scale to be specified later on, then the dispersion
relation can be re-written as
\begin{equation}
\label{disprel}
\left(\frac{\omega _{\rm phys}}{\mC}\right)^2=
\left(\frac{k_{\rm phys}}{\mC}\right)^2
-\alpha\left(\frac{k_{\rm phys}}{\mC}\right)^4 
+\beta \left(\frac{k_{\rm phys}}{\mC}\right)^6\, .
\end{equation}
It is represented in Fig.~\ref{disp}. This relation is in fact
characterized by one parameter, the ``shape parameter'' $\Upsilon $
defined by $\Upsilon \equiv 3\beta /\alpha ^2$. The derivative of the
dispersion relation vanishes at $k_{1,2}/\mC=\sqrt{\alpha /(3\beta
)}\sqrt{1\mp \sqrt{1-\Upsilon}}$ which shows that $\Upsilon <1$ and
the requirement that the dispersion relation stays positive implies
that $\Upsilon >3/4$. To summarize, one has
\begin{equation}
\frac34<\Upsilon <1\, .
\end{equation}
Obviously, the scales $k_1$ and $k_2$ only depend on the shape
of the dispersion relation, \ie only on the parameters $\alpha $ and
$\beta $. It is more convenient to express everything in terms of
$\alpha $ and $\Upsilon$. This gives
\begin{eqnarray}
\frac{k_1^2}{\mC ^2} &=& \frac{1}{\alpha \Upsilon }
\left(1-\sqrt{1-\Upsilon }\right)\, ,
\\
\frac{k_2^2}{\mC ^2} &=& \frac{1}{\alpha \Upsilon }
\left(1+\sqrt{1-\Upsilon }\right)\, .
\end{eqnarray}

In this paper, for simplicity we restrict ourselves to the case where
the background space-time is de Sitter, characterized by the constant
Hubble parameter $H_{\rm inf}$. Then, $k_0$, $\Lambda _1$ and $\Lambda
_2$ are the two scales for which $\omega _{\rm phys}=H_{\rm inf}$, see
Fig.~\ref{disp}. They depend on the parameters $\alpha $ and $\beta $
but, clearly, also on $H_{\rm inf}/m_{_{\rm C}}$. In fact, their
explicit expressions can easily be derived. For this purpose, let us
define the coefficients $Q$ and $R$ by
\begin{eqnarray}
Q &\equiv & \frac{1}{\alpha ^2\Upsilon ^2}\left(\Upsilon -1\right)\, ,
\label{defofQ} \\
R &\equiv & \frac{1}{\alpha ^3\Upsilon ^3}
\left[1-\frac32\Upsilon +\frac32\alpha \Upsilon ^2
\left(\frac{H_{\rm inf}}{m_{_{\rm C}}}\right)^2\right]\, . \label{defofR}
\end{eqnarray}
Let us notice that $H_{\rm inf}$ crosses the dispersion relation three
times only if $Q^3+R^2<0$. This implies that $H_{\rm inf}$ should be
chosen such that
\begin{equation}
H_{\rm min}<H_{\rm inf}<H_{\rm max}\, ,
\end{equation}
with $H_{\rm min}$ and $H_{\rm max}$ given by the following
expressions (which, obviously, only depend on the shape of the
dispersion relation)
\begin{eqnarray}
\frac{H_{\rm min}}{\mC} &\equiv & \frac{1}{\sqrt{\alpha
}\Upsilon}
\sqrt{\Upsilon -\frac23-\frac23\left(1-\Upsilon\right)^{3/2}}
\\
\frac{H_{\rm max}}{\mC} &\equiv &
\frac{1}{\sqrt{\alpha }\Upsilon }
\sqrt{\Upsilon-\frac23+\frac23\left(1-\Upsilon\right)^{3/2}}\, .
\end{eqnarray}
Then the three solutions can be found explicitly since they are in
fact solutions of a third order polynomial equation (more precisely,
$\omega _{\rm phys}^2=H_{_{\rm inf}}^2$ is a sixth order polynomial
equation that can be reduced to a third order equation in the variable
$k_{\rm phys}^2$). The three solutions can be written as
\begin{eqnarray}
\frac{k_0^2}{\mC ^2} &=& \frac{1}{\alpha \Upsilon }\left[1
+2\sqrt{1-\Upsilon}\cos \left(\frac{\theta +2\pi}{3}\right)\right]\, ,
\\ 
\frac{\Lambda _1^2}{\mC ^2} &=& \frac{1}{\alpha \Upsilon }\left[1
+2\sqrt{1-\Upsilon}\cos \left(\frac{\theta +4\pi}{3}\right)\right]\, ,
\\
\frac{\Lambda _2^2 }{\mC ^2} &=& \frac{1}{\alpha \Upsilon }\left[1
+2\sqrt{1-\Upsilon}\cos \left(\frac{\theta }{3}\right)\right]\, ,
\label{lambda2}
\end{eqnarray}
where $\theta \equiv \cos ^{-1}(R/\sqrt{-Q^3})$. On can check that, if
$3/4<\Upsilon<1$, then $k_0<\Lambda _1<\Lambda _2$ as required.

\par

Let us now consider a scalar field the dispersion relation of which is
given by Eq.~(\ref{disprel}). Then, as demonstrated in
Ref.~\cite{Lubo}, the vacuum expectation value 
of the energy density and pressure
are given by
\begin{widetext}
\begin{eqnarray}
\label{tmunudisp}
 \langle\rho\rangle &=& \frac{1}{4\pi^2a^4}\int _0^{+\infty}{\rm d} k k^2
     \left[a^2\left|\left(\frac{\mu_k}{a}\right)'\right|^2
     +\omega^2(k)\left|\mu_k\right|^2\right]\, ,\label{rhoeff}
\\
 \langle p\rangle&=&\frac{1}{4\pi^2a^4}\int_0^{+\infty} {\rm d}k k^2
      \left[a^2\left|\left(\frac{\mu_k}{a}\right)'\right|^2 +\left(
      \frac{2}{3}k^2\frac{{\rm d}\omega^2}{{\rm d}k^2}-\omega^2\right)
      \left|\mu_k\right|^2\right]\, .\label{peff}
\end{eqnarray}
This stress-energy tensor is conserved, \ie $\langle\rho\rangle '+
3{\cal H}\langle\rho+p\rangle=0$, with ${\cal H}\equiv a'/a$ as shown
explicitly in Ref.~\cite{LMU}. The modification of the energy density
has exactly the expected form while the modification of the pressure
is more complicated, involving the derivative of the dispersion
relation. One can easily checked that, for the linear dispersion
relation, the above formulas reduce to the standard ones.
 
\subsection{Near ultra-violet region}

Let us now try to evaluate these expressions explicitly. If we are in
a region where the WKB approximation holds, then the mode function can
be written as
\begin{equation}
\label{solgenewkb}
\mu_k(\eta)\simeq \frac{\alpha _k}{\sqrt{2\omega(k,\eta)}}\exp
\left[-i\int ^{\eta }\omega (k,\tau ){\rm d}\tau \right]+ \frac{\beta
_k}{\sqrt{2\omega(k,\eta)}}\exp \left[i\int ^{\eta } \omega (k,\tau
){\rm d}\tau \right]\, ,
\end{equation}
with $\vert\alpha(k)\vert^2 - \vert\beta(k)\vert^2=1$ from the
Wronskian normalization condition. Inserting this expression into the
formula giving the energy density, one obtains
\begin{eqnarray}
\langle \rho \rangle &=&
\frac{1}{4\pi^2a^4}\int {\rm d}k k^2\biggl\{
\frac{1}{2\omega }
\biggl[\omega ^2 +\vert \gamma \vert ^2\biggr]+
\frac{\vert \beta _k\vert ^2}{\omega }
\biggl[\omega ^2 +\vert \gamma \vert ^2\biggr]
+\frac{\alpha _k\beta _k^*}{2\omega }\biggl[\omega ^2 +\gamma ^2\biggr]
{\rm e}^{-2i\int ^{\eta }\omega (k,\tau ){\rm d}\tau }
\nonumber \\
& & +\frac{\alpha _k^*\beta _k}{2\omega }\biggl[\omega ^2 +(\gamma ^*)^2\biggr]
{\rm e}^{2i\int ^{\eta }\omega (k,\tau ){\rm d}\tau }\biggr\},
\end{eqnarray}
\end{widetext}
where we have used $\vert \alpha _k\vert ^2=1+\vert \beta _k\vert ^2$. In the 
above expression, the quantity $\gamma $ is defined as follows 
\begin{equation}\label{defg}
\gamma (k,\eta )\equiv\left[\frac{\omega'(k,\eta)}{2\omega(k,\eta)}
+i\omega(k,\eta)+\frac{a'}{a}\right]\, .
\end{equation}
In a situation where WKB is a good approximation, we have $\gamma
/\omega \simeq i$ and the previous expression reduces to
\begin{equation}
\langle \rho \rangle=\frac{1}{2\pi^2a^4}\int_{\cal K}
\dd k k^2 \left(\frac{1}{2} + \vert\beta_k\vert^2\right)\omega(k)\, ,
\label{nuven}
\end{equation}
where the domain of integration ${\cal K}$ corresponds to the region
where the WKB approximation is valid. Note that in order to remove the
two oscillatory terms, no procedure of time averaging is needed in
contrast with what was done in Ref.~\cite{Tanaka}.

\par

We now demonstrate that the energy density of Eq.~(\ref{nuven}) is
constant in time. Replacing the time-dependent comoving frequency
$\omega (k)$ by its expression in terms of the physical frequency,
namely $\omega=a\omega _{\rm phys}(k/a)$, one gets
\begin{equation}
\langle \rho \rangle=\frac{\vert\beta_k\vert^2}{2\pi^2}\int_{\rm
k_0}^{\Lambda _1} \dd k_{\rm phys} k^2_{\rm phys} \omega_{\rm
phys}(k_{\rm phys})\, ,
\label{enerphys}
\end{equation}
where we have used that $\beta _k$ is scale-independent in the case of
a de Sitter background. We have also specified the domain of
integration ${\cal K}$. It is of course crucial that this domain be
defined in terms of physical wavenumbers. As announced before,
$\langle \rho \rangle$ is time-independent. Of course, this does not
imply that the corresponding equation of state is necessarily $-1$
because, since we consider only a limited range of wavenumbers, the
energy density is not conserved. Only the total energy density,
integrated over wavenumbers from $0$ to $+\infty $, is conserved.

\par

Let us now evaluate the pressure. Repeating the same steps as before,
a long but straightforward calculation gives
\begin{widetext}
\begin{eqnarray}
\langle p \rangle &=&
\frac{1}{4\pi^2a^4}\int {\rm d}k k^2\biggl\{
\frac{1}{2\omega }
\biggl[\frac23k^2\frac{{\rm d}\omega ^2}{{\rm d}k^2}
-\omega ^2 +\vert \gamma \vert ^2\biggr]+
\frac{\vert \beta _k\vert ^2}{\omega }
\biggl[\frac23k^2\frac{{\rm d}\omega ^2}{{\rm d}k^2}
-\omega ^2 +\vert \gamma \vert ^2\biggr]
\nonumber \\
& & +\frac{\alpha _k\beta _k^*}{2\omega }\biggl[
\frac23k^2\frac{{\rm d}\omega ^2}{{\rm d}k^2}
-\omega ^2 +\gamma ^2\biggr]
{\rm e}^{-2i\int ^{\eta }\omega (k,\tau ){\rm d}\tau }
+\frac{\alpha _k^*\beta _k}{2\omega }\biggl[
\frac23k^2\frac{{\rm d}\omega ^2}{{\rm d}k^2}
-\omega ^2 +(\gamma ^*)^2\biggr]
{\rm e}^{2i\int ^{\eta }\omega (k,\tau ){\rm d}\tau }\biggr\},
\end{eqnarray}
\end{widetext}
This time, in order to remove the oscillatory terms, we can take the
time average of the previous expression, as done for the energy
density in Ref.~\cite{Tanaka}. In the following we denote the
corresponding double average by the symbol $\langle \langle \cdots
\rangle \rangle$. This yields
\begin{equation}
\langle \langle p \rangle \rangle =\frac13 
\frac{1}{2\pi^2a^4}\int_{\cal K}
\dd k k^2 \left(\frac{1}{2} + \vert\beta_k\vert^2\right)\omega(k)
\frac{{\rm d}\ln \omega ^2}{{\rm d}\ln k^2}\, . \label{nuvpr}
\end{equation}
The expression for the pressure can be evaluated explicitly. We use the
fact that the coefficient $\beta _k$ is scale-independent, thanks to
the time translation invariance. Then
\begin{equation}
\langle \langle p \rangle \rangle =\frac13
\frac{\mC^4\vert\beta_k\vert^2}{2\pi^2}
\left[{\cal P}\left(\frac{\Lambda _1}{\mC}\right)
-{\cal P}\left(\frac{k_0}{\mC}\right)\right]\, ,
\end{equation}
where the function ${\cal P}(z)$ is defined by the following
expression
\begin{eqnarray}
{\cal P}(z) &\equiv & \frac{3}{16\alpha ^2\Upsilon ^2}\frac{\omega(z)
}{z}\left[\left(9-8\Upsilon \right)+2\alpha \Upsilon z^2+\frac83\alpha
^2\Upsilon ^2z^4\right] 
\nonumber \\
&+& \frac{27}{32\alpha ^2\Upsilon
^2}\sqrt{\frac{3}{\Upsilon }} \left(1-\frac43\Upsilon \right)\ln
\biggl[\sqrt{\frac{3}{\Upsilon }} \left(\frac23\alpha \Upsilon
z^2-1\right)
\nonumber \\
&+& 2\frac{\omega(z)}{z}\biggr]\, ,
\end{eqnarray}
with $\omega (z)\equiv \sqrt{z^2-\alpha z^4+\alpha ^2\Upsilon z^6/3}$.

\par

\begin{figure*}
\includegraphics[width=.495\textwidth,height=.4\textwidth]{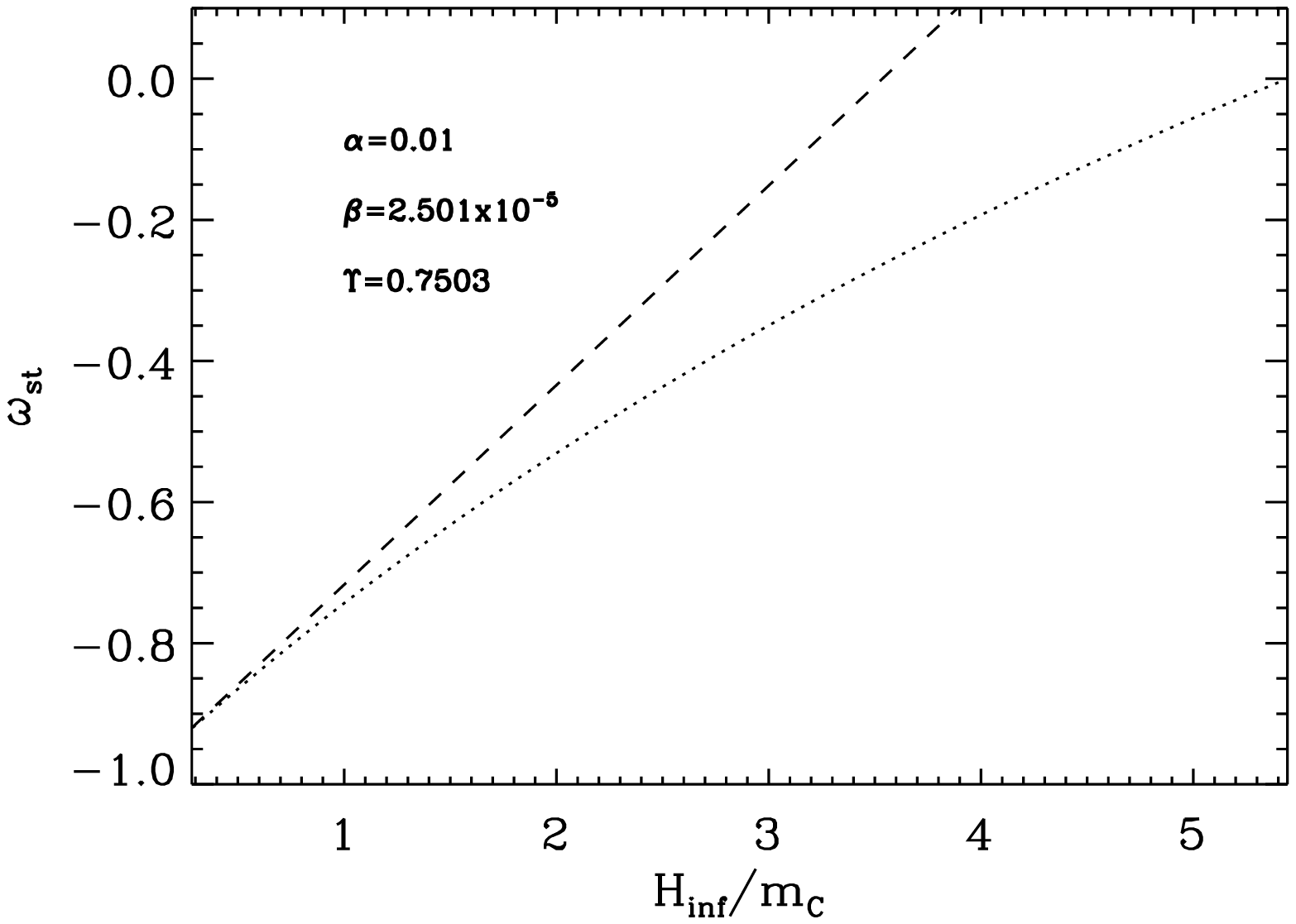}
\includegraphics[width=.495\textwidth,height=.4\textwidth]{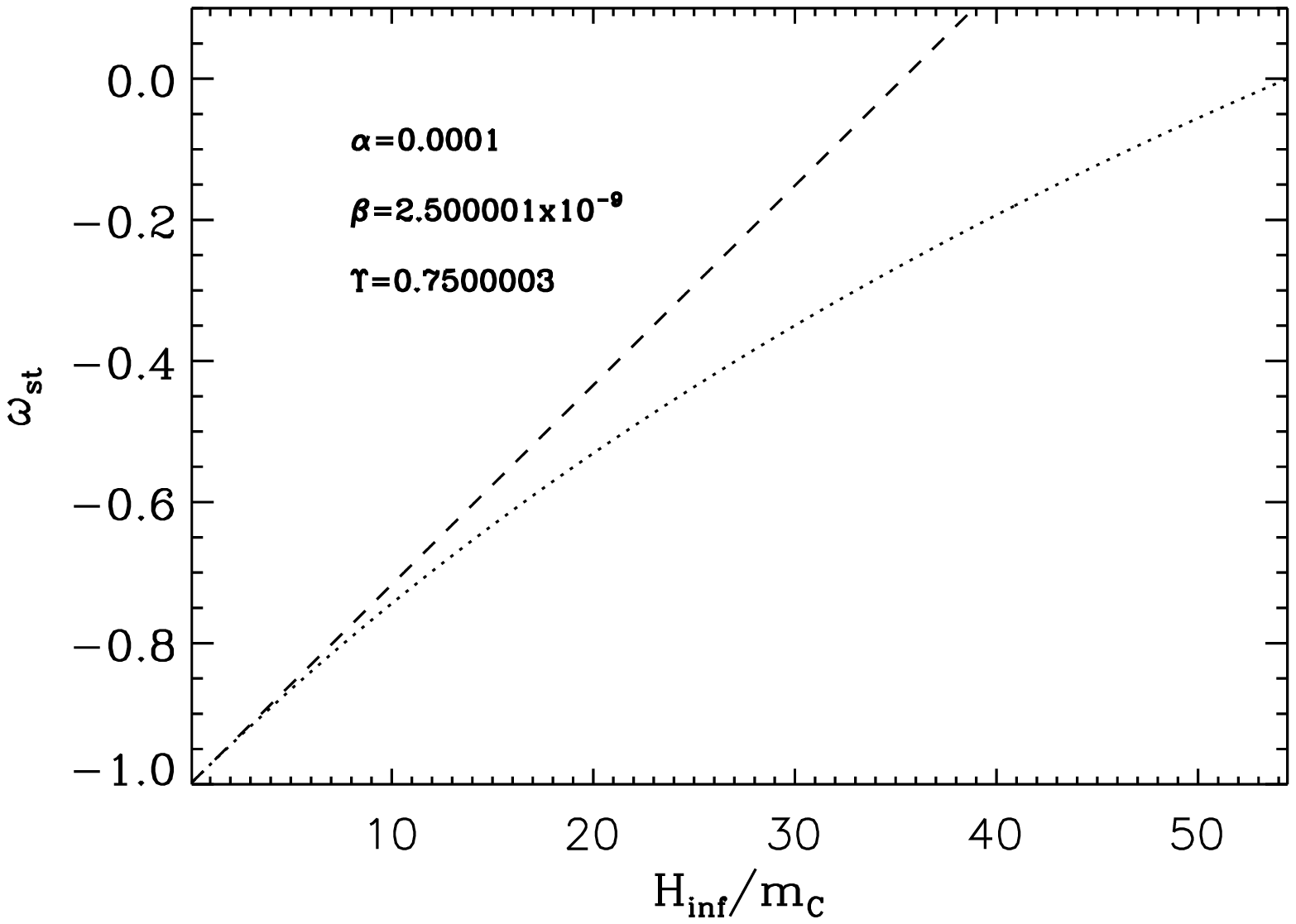}
\caption{Left panel: Equation of state versus $H_{\rm inf}/\mC$ for
$\alpha =0.01$ and $\Upsilon=0.7003$ (dotted line) computed according
to Eq.~(\ref{statenearuv}). The dashed line is an approximation of the
equation of state, valid for small values of $H_{\rm inf}/\mC$, and
derived in Eq.~(\ref{statenearuvapprox}). Right panel: same as left
panel but with $\alpha =0.0001$ and $\Upsilon=0.700003$.}
\label{state_nearuv}
\end{figure*}

{F}rom the above expression, the energy density can be evaluated very
simply if one notices that
\begin{equation}
\int _{\cal K}{\rm d}kk^2\omega (k)\frac{{\rm d}\ln \omega ^2}{{\rm
d}\ln k^2} =k^3\omega (k)\vert _{\cal K}-3\int _{\cal K}{\rm
d}kk^2\omega (k)\, .
\end{equation}
Then the equation of state $\omega _{\rm st}\equiv p/\rho $ can be
expressed as 
\begin{equation}
\label{statenearuv}
\omega _{\rm st}=\left[\left(\frac{H_{\rm
inf}}{\mC}\right)\frac{(\Lambda _1/\mC)^3-(k_0/\mC)^3}{ {\cal
P}\left(\Lambda _1/\mC\right) -{\cal
P}\left(k_0/\mC\right)}-1\right]^{-1}\, .
\end{equation}
A typical example is represented in Fig.~\ref{state_nearuv}. The
striking feature of this plot is that, for $H_{_{\rm inf}}/\mC \ll 1$,
the equation of state goes to one. Let us emphasize that this regime
corresponds to a physical requirement that should be met if one wants
to be in a realistic situation. In order to understand better how this
happens, we perform the following perturbative treatment.

\par

If the Hubble parameter is small in comparison with the new scale
$\mC$, then the minimum of the dispersion relation is close to zero
which in turn means that $\Upsilon \simeq 3/4$. Admittedly, this is a
fine-tuning of the shape of the dispersion relation. The above
considerations suggest that that Taylor expansion in $\Upsilon -3/4$
can be performed. Then, one finds [see Eqs.~(\ref{defofQ})
and~(\ref{defofR})]
\begin{equation}
\label{dlRQ}
\frac{R}{\sqrt{-Q^3}}=-1+\frac{27}{4}\alpha \left(\frac{H_{\rm
inf}}{\mC}\right)^2+{\cal O}\left(\Upsilon -\frac34\right)\, .
\end{equation}
The next step is to expand the $\cos ^{-1}$ function which appears in
the paragraph following (\ref{lambda2}) with the above
argument. At this point, since we expect
$H_{\rm inf} \ll m_c$, one can treat
$\alpha (H_{\rm inf}/\mC)^2$ as a small parameter and Taylor expand
our expressions, e.g. Eq.~(\ref{dlRQ}), in this parameter.
However, this
approximation will break down when $H_{\rm inf}$ becomes large,
especially far from $H_{\rm min}$. This is true even if $\Upsilon=3/4$
since then $\alpha (H_{\rm max}/\mC)^2 =8/27\simeq 0.3$ which
represents an error of $\simeq 30\%$. In the vicinity of $H_{\rm min}$
the approximation is of course much better. Then, working at zeroth
order in $\Upsilon -3/4$, it is easy to show that the angle $\theta$
introduced in the paragraph after (\ref{lambda2}) is
\begin{eqnarray}
\theta &=& \pi -\sqrt{\frac{27}{2}\alpha \left(\frac{H_{\rm
inf}}{\mC}\right)^2}+{\cal O}\left\{\left[\alpha \left(\frac{H_{\rm
inf}}{\mC}\right)^2\right]^{3/2}\right\}
\nonumber \\
& & +{\cal O}\left(\Upsilon -\frac34\right)
\, .
\end{eqnarray}
If one inserts the above expression into the formulas giving $k_0$ and
$\Lambda _1$, one obtains
\begin{widetext}
\begin{eqnarray}
\left(\frac{k_0}{\mC }\right)^2 &=& \frac{4}{3\alpha
}\left\{\frac34\alpha \left(\frac{H_{\rm inf}}{\mC}\right)^2+{\cal
O}\left[\alpha ^2\left(\frac{H_{\rm inf}}{\mC}\right)^4\right]
\right\}+{\cal O}\left(\Upsilon -\frac34\right)\, ,
\\ 
\left(\frac{\Lambda _1}{\mC }\right)^2
&=& \frac{2}{\alpha
}\left\{1-\sqrt{\frac{\alpha}{2} \left(\frac{H_{\rm
inf}}{\mC}\right)^2}-\frac{\alpha}{4}
\left(\frac{H_{\rm inf}}{\mC}\right)^2
+{\cal
O}\left[\alpha ^{3/2}\left(\frac{H_{\rm inf}}{\mC}\right)^3\right]
\right\}+{\cal O}\left(\Upsilon -\frac34\right)\, .
\end{eqnarray}
The first expression is expected since it says that at leading order
$k_0\simeq H_{\rm inf}$. This is because the usual transition between
sub and super-Hubble modes occurs in the region where the dispersion
relation is almost linear. In the same manner, one has
\begin{eqnarray}
{\cal P}\left(\frac{k_0}{\mC }\right) &=& \frac{1}{\alpha ^2}
\left\{1+\frac12\alpha \left(\frac{H_{\rm inf}}{\mC}\right)^2 +{\cal
O}\left[\alpha ^2\left(\frac{H_{\rm inf}}{\mC}\right)^4\right]
\right\}+{\cal O}\left(\Upsilon -\frac34\right)\, ,
\\ 
{\cal P}\left(\frac{\Lambda _1}{\mC }\right) &=& \frac{1}{\alpha ^2
}\left(\frac{H_{\rm inf}}{\Lambda _1}\right)
\left\{4-5\sqrt{\frac{\alpha }{2}\left(\frac{H_{\rm
inf}}{\mC}\right)^2} -\frac{1}{4}\alpha \left(\frac{H_{\rm
inf}}{\mC}\right)^2 +{\cal O}\left[\alpha ^{3/2}\left(\frac{H_{\rm
inf}}{\mC}\right)^3\right] \right\}+{\cal O}\left(\Upsilon
-\frac34\right)\, .
\end{eqnarray}
\end{widetext}
Finally, putting everything together, at leading order in $\Upsilon
-3/4$ and in $\alpha (H_{\rm inf}/\mC)^2$, we obtain a simple equation
in the regime of interest, namely
\begin{equation}
\label{statenearuvapprox}
\omega _{\rm st}\simeq -1+2\sqrt{2\alpha \left(\frac{H_{\rm
inf}}{\mC}\right)^2}\, .
\end{equation}
It is represented by a dashed line in Fig.~\ref{state_nearuv}. We are
now in a position where the behavior of the equation of state can be
understood better. In the limit $\Upsilon \rightarrow 3/4$ and $H_{\rm
inf}\rightarrow H_{\rm min}$, the equation of state goes to $-1$. Let
us notice that, since we have $H_{\rm min}^2=8(\Upsilon -3/4)/(3\alpha
)+\cdots $, the limit $H\rightarrow H_{\rm min}$ corresponds, in this
case, to $H_{\rm inf}\rightarrow 0$. We conclude that, in the physical
regime of interest, namely $H/\mC\ll 1$ the equation of state is
extremely close to that of the vacuum. Far from $H_{\rm min}$ the
approximation used above breaks down as is apparent from
Fig.~\ref{state_nearuv}. When $H_{\rm inf}\rightarrow H_{\rm max}$, it
is clear that $k_0\rightarrow \Lambda _1$ and the expression giving
the equation of state becomes ambiguous. From the plots, we see that
$\omega _{\rm st}$ goes in fact to zero. Therefore, even if the Hubble
constant is not small in comparison with $\mC$, the equation of state
$\omega _{\rm st}$ remains negative.

\subsection{Far ultra-violet region}

Let us now consider the far ultra-violet region. We fix the initial
conditions in the region where the WKB approximation holds by
selecting positive frequency modes, which corresponds to the choice of
the Bunch-Davies adiabatic vacuum state for the field,
\begin{equation}
\mu_k(\eta)=\frac{1}{\sqrt{2\omega(k,\eta)}}\exp \left[-i \int
^\eta\omega(k,\tau )\dd\tau \right]\, .
\end{equation}
At some time $\eta_2(k)$, the comoving mode enters the far
ultra-violet region, \ie when the physical wavelength equal $\Lambda
_2$, and $\mu_k$ and its first derivative must be matched to the above
solution, which gives
\begin{eqnarray}
\mu_k(\eta)&=&\frac{1}{\sqrt{2\omega[k,\eta_2(k)]}}\hbox{e}^{-i
\int_{\eta_{\rm i}}^{\eta_2(k)}\omega(k,\tau )\dd\tau }
\frac{a(\eta)}{a(\eta_2)}\nonumber \\
& \times & \left\{1-\gamma (k,\eta
_1)\int_{\eta_2(k)}^\eta\left[\frac{a(\eta_2)}{a(\tau)}
\right]^2\dd\tau \right\} ,
\end{eqnarray}
where the quantity $\gamma$ has been defined previously, see
Eq.~(\ref{defg}). Then, one can safely neglect the second term which
is the decaying mode and express everything in terms of physical
quantities, using that the scale factor can be written as $a(\eta
)=-1/(H_{\rm inf}\eta )$. This gives
\begin{equation}
\vert \mu_k(\eta)\vert ^2 \simeq \frac{1}{\omega[k,\eta_2(k)]}
\frac{\Lambda _2^2}{k_{\rm phys}^2(\eta )}\, .
\end{equation}
But, one has $\omega[k,\eta_2(k)]=a(\eta _2)\omega _{\rm
phys}[k/a(\eta _2)]=a(\eta _2)\omega _{\rm phys}(\Lambda _2) =a(\eta
_2)H_{\rm inf}$. Since $a[\eta _2(k)]=k/\Lambda _2$, one finally
arrives at
\begin{equation}
\label{modefaruv}
\vert \mu_k(\eta)\vert ^2 =\frac{1}{2k}
\frac{\Lambda _2^3}{H_{\rm inf}k_{\rm phys}^2(\eta )}\, .
\end{equation}
Then, it is straightforward to calculate the energy density and the
pressure. This gives
\begin{widetext}
\begin{eqnarray}
\langle \rho\rangle &=& \frac{1}{16\pi ^2}\frac{\Lambda
_2^3\mC^2}{H_{\rm inf}} \left\{\left[\left(\frac{\Lambda
_2}{\mC}\right)^2-\left(\frac{\Lambda _1}{\mC}\right)^2\right]
-\frac{\alpha
}{2}\left[ \left(\frac{\Lambda
_2}{\mC}\right)^4-\left(\frac{\Lambda _1}{\mC}\right)^4\right]
+\frac{\beta}{3}\left[ \left(\frac{\Lambda
_2}{\mC}\right)^6-\left(\frac{\Lambda _1}{\mC}\right)^6\right]\right\}\, ,
\label{fuven}
\\
\langle p\rangle &=& -\frac{1}{48\pi ^2}\frac{\Lambda
_2^3\mC^2}{H_{\rm inf}} \left\{\left[\left(\frac{\Lambda
_2}{\mC}\right)^2-\left(\frac{\Lambda _1}{\mC}\right)^2\right]
+\frac{\alpha
}{2}\left[ \left(\frac{\Lambda
_2}{\mC}\right)^4-\left(\frac{\Lambda _1}{\mC}\right)^4\right]
-\beta\left[ \left(\frac{\Lambda
_2}{\mC}\right)^6-\left(\frac{\Lambda _1}{\mC}\right)^6\right]\right\}
\, . \label{fuvpr}
\end{eqnarray}
In the standard case where $\alpha =\beta =0$, one recovers that
$p/\rho=-1/3$ as required for super-horizon modes (see e.g. 
\cite{ABM}). Then,
straightforward algebraic manipulations show that
\begin{eqnarray}
\label{linkprhofaruv}
\langle \rho \rangle &=& -\langle p\rangle +\frac23\frac{\Lambda
_2^3\mC^2}{H_{\rm inf}}\left\{\left[
\left(\frac{\Lambda
_2}{\mC}\right)^2-\left(\frac{\Lambda _1}{\mC}\right)^2\right]
-\alpha \left[\left(\frac{\Lambda
_2}{\mC}\right)^4-\left(\frac{\Lambda _1}{\mC}\right)^4\right]
+\beta \left[\left(\frac{\Lambda
_2}{\mC}\right)^6-\left(\frac{\Lambda _1}{\mC}\right)^6\right]
\right\}
\\
&=& -\langle p\rangle +\frac23\frac{\Lambda
_2^3\mC^2}{H_{\rm inf}}\left[\omega _{\rm phys}^2\left(\Lambda _2\right)-
\omega _{\rm phys}^2\left(\Lambda _1\right)\right]=-\langle p\rangle \, .
\end{eqnarray}
\end{widetext}
Therefore, in the far ultra-violet region, the equation of state is
nothing but the vacuum equation of state, \ie $-1$.

\par

Can we understand this result better? For this purpose let us consider
Eqs.~(\ref{tmunudisp}) again. The terms $(\mu _k/a)'$ vanish because
$\mu _k\propto a$ in the region under consideration. Then, the link
between the pressure and the energy density can be re-written as
\begin{equation}
\langle p\rangle=-\langle \rho \rangle +\frac23\frac{1}{4\pi
^2a^4}\int _{\cal K}{\rm d}kk^4 \frac{{\rm d}\omega ^2}{{\rm
d}k^2}\vert \mu _k\vert ^2\, .
\end{equation}
Now, if $\vert \mu _k\vert ^2$ scales as $\vert \mu _k\vert ^2 \propto
1/k^3$ as indicated by Eq.~(\ref{modefaruv}) which is the consequence
of having matched the mode function in the far ultra-violet region to
the initial Bunch-Davis vacuum, then the above expression can be
re-written as 
\begin{equation}
\langle p\rangle+\langle \rho \rangle \propto \frac23 \int _{\cal
K}{\rm d}k \frac{{\rm d}\omega ^2}{{\rm d}k}\, .
\end{equation}
Clearly, if the interval ${\cal K}$ is such that the frequency is the
same at its boundaries then we obtain the equation of state of the
vacuum. This conclusion does not depend on the detailed shape of the
dispersion relation in this region. It is also obvious that the
calculations done previously for a specific dispersion relation are
fully compatible with the above considerations, in particular the
difference between $\langle p\rangle$ and -$\langle \rho \rangle$ in
Eq.~(\ref{linkprhofaruv}) is given by a term equal to $2/3$ times the
difference between the square of the effective frequency at the
boundaries of the far ultra-violet region.

\par

To conclude this section, one can estimate the equation of state of
the created particles coming from the whole ultra-violet region. In
the near ultra-violet region, one has
\begin{equation}
\label{puvnear}
\langle p\rangle _{\rm near-UV}=\left[-1+{\cal O}\left(\frac{H_{\rm
inf}}{\mC}\right)\right]\langle \rho \rangle _{\rm near-UV}\, ,
\end{equation}
while in the far ultra-violet region 
\begin{equation}
\label{puvfar}
\langle p\rangle _{\rm far-UV}=-\langle \rho \rangle _{\rm far-UV}\, .
\end{equation}
In the ``ultra-far'' region ($k_{\rm phys}>\Lambda _2$), \ie in the
region where the initial condition are fixed, we have by definition
the adiabatic vacuum and, therefore, no created particles. Hence, we
do not need to take into account this region. As a consequence, the
equation of state of the whole ultra-violet region can be written as
\begin{equation}
\omega _{\rm st-UV}\equiv \frac{\langle p\rangle _{\rm near-UV}
+\langle p\rangle _{\rm far-UV}}{\langle \rho \rangle _{\rm near-UV}
+\langle \rho \rangle _{\rm far-UV}}\simeq -1+{\cal
O}\left(\frac{H_{\rm inf}}{\mC}\right)\, ,
\end{equation}
where we have used Eqs.~(\ref{puvnear}) and~(\ref{puvfar}). Since the 
total energy density 
\begin{equation}
\label{totalrho}
\langle \rho \rangle _{_{\rm UV}}\equiv \langle \rho \rangle _{\rm
near-UV} +\langle \rho \rangle _{\rm far-UV}\, ,
\end{equation}
is constant in time, we see that the energy density of the created
particles almost behave as a positive cosmological constant. The
slight non-conservation is due to the fact that we have not taken into
account the infra-red region which also contributes (let us recall
that the total energy-momentum tensor is conserved exactly). The same
calculation performed with a linear dispersion relation would have led
to the result $\omega _{\rm st-UV}=1/3$. We have thus demonstrated
explicitly that taking into account the trans-Planckian corrections is
important when one evaluates the back-reaction. 

\par

We end this section with the following remark. It is important to keep
in mind that the previous calculation is not the calculation of the
equation of state cosmological perturbations. Therefore, one cannot
claim that the equation of state of the cosmological perturbations
with a modified dispersion relation is the one of a cosmological
constant. What has been calculated is just the equation of state of a
test field. Nevertheless, we have shown with the help of the previous
toy model that any attempt to evaluate the equation of state of the
cosmological perturbations in a regime where the dispersion relation
is modified must take into account these trans-Planckian corrections.

\section{Discussion and Conclusions}

Now that we have determined the energy density and pressure of the
ultraviolet modes, we will study their back-reaction on the background
space-time and matter. We work in the context of the toy model studied
in the previous section. We assume that at early times all modes with
wavenumber larger than $\Lambda_2$ start out in their adiabatic
vacuum. Thus, after subtraction of the quantum vacuum terms, these
modes do not contribute to the energy-momentum tensor, and thus there
will be no remaining ultraviolet divergences.

\par

We now follow a mode with fixed comoving wavelength which starts out
deep in the ultraviolet region ($k_{\rm phys}>\Lambda_2$) in its
adiabatic vacuum. The mode will then spend a finite time interval in
the intermediate frequency range $\Lambda_1<k_{\rm phys} < \Lambda_2$
during which the adiabaticity condition for mode evolution is
violated, the state gets squeezed, and, from the point of view of the
vacuum state for $k_{\rm phys}<\Lambda_1$, a non-vanishing occupation
number $n_k$ is generated. This occupation number remains constant
when $k_0<k_{\rm phys} < \Lambda_1$, since in this frequency range the
adiabaticity condition is restored.

\par

Given this setup, we consider the energy density $\rho_{UV}$ of
ultraviolet modes (modes with wavelength smaller than the Hubble
radius) which is the sum of~(\ref{nuven}) and~(\ref{fuven}), see
Eq.~(\ref{totalrho}). By time-translation invariance of the
background, both terms are independent of time in a de Sitter
background in which $H_{\rm inf}$ is constant. Thus,
\begin{equation} 
\label{constancy}
\frac{{\rm d}\rho_{_{\rm UV}}}{{\rm d}t} \, = \, 0 \, .
\end{equation}
This seems to indicate that the ultraviolet energy density evolves
like a cosmological constant, and its back-reaction, rather than
preventing inflation, will simply lead to a renormalization of the
cosmological constant.

\par

Naive intuition, namely treating the equation of state as radiative,
\ie $p_{_{\rm UV}}=\rho_{_{\rm UV}}/3$, would have led to a problem
with this conclusion, namely an extreme non-conservation of the
energy-momentum tensor of the ultraviolet modes, a non-conservation on
the energy density scale of $\Lambda _1^4$ (in this context the
difference between $\Lambda_1$ and $\Lambda_2$ is not
relevant). However, our analysis of the previous section has shown
that in fact the energy-momentum tensor of the ultraviolet modes is
that of a cosmological constant, up to correction terms which are
suppressed by a factor of $H_{\rm inf}/\mC$ compared to the dominant
terms.

\par
 
Let us take a look at the equations of back-reaction. Following the
method discussed in Refs.~\cite{Abramo1,Abramo2} in the context of the
problem of the back-reaction of infrared cosmological
fluctuations. The back-reaction effect of interest here is the fact
that linear cosmological fluctuations effect the background metric and
matter if one works out the Einstein equations to quadratic order in
the amplitude of the primordial perturbations. To be specific, in the
following we shall consider a homogeneous, isotropic and spatially
flat background, and will take matter to be a scalar field $\varphi$
with a quadratic potential given by the scalar field mass $m$. For the
fluctuations we parameterize the equation of state as
\begin{equation} \label{eos2}
p_{_{\rm UV}}\equiv (-1+\alpha )\rho_{_{\rm UV}} \, ,
\end{equation}
where (by the above discussion) $\alpha $ is a positive constant
expected to be of the order $H_{\rm inf}/\mC$.

\par

Our starting point consists of taking the FLRW equations
\begin{eqnarray}
\label{FRW}
H^2 &=& {\kappa \over 3} \rho \, , \quad
\frac{{\rm d}\rho}{{\rm d}t} = - 3 H (\rho + p) \, , 
\end{eqnarray}
where $\kappa \equiv 8\pi /\mP^2$. The presence of the fluctuations
produces a back-reaction effect on the background which is quadratic
in the amplitude of the fluctuations~\cite{MAB,ABM} and leads to a
correction of both metric and matter, \ie to corrections $\delta H$ of
the Hubble expansion rate and $\delta \varphi $ of the scalar field
\begin{eqnarray}
 \label{bransatz}
H= H_{\rm inf}+ \delta H \, ,\quad 
\varphi = \varphi_{\rm inf} + \delta \varphi  \, , 
\end{eqnarray}
where the subscripts ``${\rm inf}$'' stand for the quantities
evaluated in the unperturbed background. The back-reaction of the
linear fluctuations on the background is described by contributions
$\rho_{\rm br}=\rho _{_{\rm UV}}$ and $p_{\rm br}=p_{_{\rm UV}}$ (and
we use the ultraviolet energy densities and pressures as the
back-reaction quantities) to the energy density and pressure. If we
insert the back-reaction ansatz~(\ref{bransatz}) into the
equations~(\ref{FRW}) (written in the slow-roll approximation) and
linearize in $\delta H$ and $\delta \varphi $, we obtain the following
equations
\begin{widetext}
\begin{eqnarray}
\label{lin1} 
2 H_{\rm inf}\, \delta H &=& {\kappa \over 3}\left(
\rho_{_{\rm UV}} + m^2 \varphi_{\rm inf}\delta \varphi \right)\,
,\quad m^2\left(\frac{{\rm d}\varphi_{\rm inf}}{{\rm d}t}\right)\delta
\varphi +m^2\varphi_{\rm inf}\left(\frac{{\rm d}\delta \varphi }{{\rm
d}t}\right) = -3 H_{\rm inf}\left( \rho_{_{\rm UV}} + p_{_{\rm UV}}
\right) -\frac{{\rm d}\rho_{_{\rm UV}}}{{\rm d}t} \, .
\end{eqnarray}
\end{widetext}
We have assumed that the background model has a massive potential, \ie
$V(\varphi)=m^2\varphi ^2/2$. The back-reaction of infrared modes
(modes with wavelength larger than the Hubble radius) was analyzed in
\cite{MAB,ABM} (see also Ref.~\cite{TW}) and was shown to correspond
to a negative cosmological constant whose absolute value increases in
time (see Ref.~\cite{RBrev2} for resulting speculations on how this
effect might be used to address the cosmological constant
problem). Here, we will study the back-reaction effects of the
ultraviolet modes for which ${\rm d}\rho_{\rm br}/{\rm d}t = {\rm
d}\rho_{_{\rm UV }}/{\rm d}t =0$ and for which the equation of state
is given by (\ref{eos2}). In this case, the linearized Klein-Gordon
equation can be simplified to yield
\begin{equation} 
\label{lin3}
\frac{{\rm d}\delta \varphi }{{\rm d}t}+\frac{1}{\varphi _{\rm
inf}}\left( \frac{{\rm d}\varphi_{\rm inf}}{{\rm d}t}\right) \delta
\varphi = - {{3 \alpha H_{\rm inf}} \over {m^2 \varphi_{\rm inf}}} 
\rho_{_{\rm UV}} \, .
\end{equation}
The solution of Eqs.~(\ref{lin3}) and (\ref{lin1}) is
\begin{eqnarray}
\delta \varphi &=&  
-3\sqrt{\frac{\kappa }{6}}\left(\frac{\rho _{_{\rm UV}}}{m}\right)
\alpha t \label{br1} \, ,\\
\delta H &=& \frac{\kappa }{6} \frac{\rho_{_{\rm UV}}}{H_{\rm inf}} 
\left(1 - 3\alpha H_{\rm inf}t\right) \, .
\label{br2}
\end{eqnarray}
As expected $\delta \varphi $ and $\delta H$ vanish if $\rho _{_{\rm
UV}}=0$. As stressed in \cite{Unruh} and later analyzed in detail in
\cite{Ghazal}, it is important to express the result in terms of
physical observables instead of in terms of the non-measurable
background time coordinate $t$.  The obvious clock in our simple
system is the scalar field $\varphi \equiv \varphi _{\rm inf}+\delta
\varphi $ itself.  After a simple calculation it follows that the
back-reaction effect $\delta H$ measured in terms of $\varphi$ is
given by
\begin{equation}
\label{br3}
H_{\rm inf}+\delta H=\sqrt{\frac{\kappa }{6}}
\left(m\varphi +\sqrt{\frac{\kappa }{6}}
\frac{\rho _{_{\rm UV}}}{H_{\rm inf}}\right)\, .
\end{equation} 
Thus, the effect of the ultraviolet back-reaction terms in a de Sitter
background corresponds to a positive re-normalization of the cosmological
constant. This result does not depend on the value of $\alpha$.

\par

So far, we have shown that a large ultraviolet back-reaction does not
prevent inflation, in contrast to naive expectations. Instead, it
leads to a re-normalization of the cosmological constant. On the other
hand, as is evident from Eq.~(\ref{br2}), the back-reaction terms can
lead to a faster rolling of the scalar field $\varphi_{\rm inf}$.
However, for the small values of $\alpha $ of the order of $H_{\rm
inf}/\mC$ indicated by our analysis, the increase in the rolling speed
does not prevent a phase of inflation of sufficient length (\ie having
$|\delta \varphi | < |\varphi _{\rm inf}|$ for a time interval
$t\simeq H_{\rm inf}^{-1}$) as long as
\begin{equation}
n_k \lsim {H _{\rm inf}\over \mP} \, ,
\end{equation}
which is a less stringent constraint than the one derived by Tanaka,
and still allows observable effects on cosmic microwave fluctuations
from trans-Planckian physics.


We have presented an attempt to study the effects of back-reaction on
the trans-Planckian problem of inflationary cosmology. The question
initially posed in Refs.~\cite{Tanaka} and~\cite{Starob} is whether
the back-reaction of a state which consists of excited modes during
the inflationary phase on scales smaller than the Hubble radius will
prevent inflation. We have seen that the analysis is more subtle than
it initially appears from the above works. We have shown that the
back-reaction of an excited state preserving time-translation
invariance does not prevent inflation, but simply leads to a
renormalization of the Hubble constant. However, back-reaction will
lead to a slightly faster rolling of the scalar field. As long as the
occupation numbers $n_k$ are smaller than $H_{\rm inf}/\mP$, the
rolling will be consistent with the standard inflationary
paradigm. Such occupation numbers can lead to observable effects on
the cosmic microwave background.

\par

To render the analysis well defined, we have considered a dispersion
relation for which the violation of adiabaticity is concentrated in a
finite range of physical wavenumbers $\Lambda _1< k_{\rm phys} <
\Lambda _2$. We have assumed that all modes start off in their
adiabatic vacuum state when $k_{\rm phys} > \Lambda_2$. They are
squeezed while $\Lambda_1< k_{\rm phys} < \Lambda _2$, and then emerge
as excited states when $k_{\rm phys} < \Lambda_1$. Let us also notice
that by making the wavelength interval $\Lambda _1< k_{\rm phys} <
\Lambda _2$ small, the energy density in the far ultraviolet modes can
always be made small compared to the cutoff energy density $\Lambda
_1^4$ (or $\Lambda _2^4$). Thus, in our approach the Planck energy
density problem recently discussed in Ref.~\cite{Jacob} in an approach
to quantum field theory on a growing lattice, in which the number of
fundamental field modes is increasing in time, does not arise (the
analysis of Ref.~\cite{Jacob} finds that the continual creation of
modes at the cutoff scale yields an energy density which is of Planck
scale). In our analysis, the Hilbert space of modes is
time-independent, but as time proceeds an increasing subset of this
space gets populated. Our mechanism for continual excitation of new
modes appears to be more smooth than the lattice approach of
Ref.~\cite{Jacob}.
 
\par

There are several important deficiencies in our analysis. First, we
have used an ad-hoc regularization and renormalization prescription
which consists of imposing effectively an abrupt cutoff in momentum
space, and of subtracting the ground state energy of each field
Fourier mode. This procedure is not covariant. It would be of interest
to study our problem using a mathematically more rigorous
regularization prescription, such as adiabatic regularization (see
e.g. Ref.~\cite{Gianni} and references therein).

\par

Another serious concern is that we have considered matter fluctuations
without taking into account the induced metric fluctuations. It is
well known that the inclusion of metric fluctuations leads to
dramatically different results for super-Hubble-scale
perturbations~\cite{MAB,ABM}. Thus, one might expect that the
gravitational fluctuations could play an important role in the far
ultraviolet region where $\omega_{\rm phys} < H_{\rm inf}$. However,
at the present time we are not able to study this issue because the
non-standard dispersion relation has been set up for the matter sector
only.  It is a challenge for future research to include the presence
of non-standard dispersion relations consistently in both the
gravitational and matter sector.

\vskip0.5cm
\centerline{\bf Acknowledgments}

We wish to thank Ted Jacobson and Martin Lemoine for interesting
discussions. During the course of this work, R.~B. was supported by
the Perimeter Institute for Theoretical Physics, and (at Brown) by the
US Department of Energy under Contract DE-FG02-91ER40688,
TASK~A. J.~M.  thanks the High Energy Group of Brown University and
the Perimeter Institute for Theoretical Physics for warm hospitality.

\end{document}